**Title:** Illustrating the structures of bias from immortal time using directed acyclic graphs

**Short title:** Bias from immortal time

**Authors**:

Guoyi Yang[1], Stephen Burgess[2,3], C Mary Schooling[1,4*]

[1] School of Public Health, Li Ka Shing Faculty of Medicine, The University of Hong Kong, Hong Kong, China

[2] MRC Biostatistics Unit, University of Cambridge, Cambridge, UK

[3] British Heart Foundation Cardiovascular Epidemiology Unit, Department of Public Health and Primary Care, University of Cambridge, Cambridge, UK

[4] Graduate School of Public Health and Health Policy, City University of New York, New York, United States

[*]**Corresponding Author:**

C Mary Schooling

School of Public Health, Li Ka Shing Faculty of Medicine, The University of Hong Kong.

Email address: cms1@hku.hk.

**Word count:** 2917

**Abstract**

**Background:** Immortal time is a period of follow-up during which death or the study outcome cannot occur by design. Bias from immortal time has been increasingly recognized in epidemiologic studies. However, the fundamental causes and structures of bias from immortal time have not been explained systematically using a structural approach.

**Methods:** We use an example "Do Nobel Prize winners live longer than less recognized scientists?" for illustration. We illustrate how immortal time arises and present the structures of bias from immortal time using time-varying directed acyclic graphs (DAGs). We further explore the structures of bias with the exclusion of immortal time and with the presence of competing risks. We discuss how these structures are shared by different study designs in pharmacoepidemiology and provide solutions, where possible, to address the bias.

**Results:** We illustrate that immortal time arises from using postbaseline information to define exposure or eligibility. We use time-varying DAGs to explain the structures of bias from immortal time are confounding by survival until exposure allocation or selection bias from selecting on survival until eligibility. We explain that excluding immortal time from the follow-up does not fully address this confounding or selection bias, and that the presence of competing risks can worsen the bias. Bias from immortal time may be avoided by aligning time zero, exposure allocation and eligibility, and by excluding individuals with prior exposure.

**Conclusions:** Understanding bias from immortal time in terms of confounding or selection bias helps researchers identify and thereby avoid or ameliorate this bias.

**Key messages:**

- Immortal time arises from using postbaseline information to define exposure or eligibility.

- Using time-varying directed acyclic graphs reveals that the structures of bias from immortal time are confounding by survival until exposure allocation or selection bias from selecting on survival until eligibility.

- Excluding immortal time from the follow-up does not fully address bias from immortal time, and the presence of competing risks can worsen the bias.

**Introduction**

Immortal time refers to a period of follow-up during which death or the study outcome cannot occur by design.[1] Bias from immortal time was first identified in the 1970s,[2, 3] and has been described as a bias resulting from counting follow-up times incorrectly in terms of exposure status.[4, 5] Although bias from immortal time has been warned against for decades, it is still increasingly evident in epidemiologic studies.[6-12] This is possibly because the fundamental causes and structures of bias from immortal time have not been explained systematically and comprehensively using a structural approach.

Directed acyclic graphs (DAGs) are useful in illustrating causal structures and thereby preventing the key sources of bias in epidemiologic studies, i.e., confounding (the existence of common causes of exposure and outcome) and selection bias (conditioning on common consequences of exposure and outcome).[13-15] Previous DAGs for bias from immortal time do not reflect its time-varying nature,[16] or take into account different study designs.[17, 18] Time-varying DAGs include separate nodes for a variable at different times,[19] which can depict time-varying exposures accurately.

Here, we illustrate how immortal time arises and present the structures of bias from immortal time as confounding or selection bias using time-varying DAGs. We further explore the structures of bias with the exclusion of immortal time and with the presence of competing risks. We then discuss how these structures are shared by different study designs in pharmacoepidemiology and provide solutions, where possible, to address the bias.

**Do Nobel Prize winners live longer than less recognized scientists?**

We use an example "Do Nobel Prize winners live longer than less recognized scientists?" for illustration, because there is a time lag between the publication of a scientific discovery and the conferment of a Nobel Prize.[20] Consider a study to investigate the survival benefit of winning a Nobel Prize. All scientists who won at least one Nobel Prize were identified as Nobel Prize winners. For each winner, a control was selected as a scientist who was the same sex, was born in the same era, and worked in the same institution as the winner when the discovery was published. For simplicity, we suppose there are no other confounders.

For illustration, we use time-varying DAGs with two time points (0 and 1) and follow-up continuing beyond time 1, as previously.[17] $E_0$ and $E_1$ denote exposure status at time 0 and 1, respectively. $D_{0+}$ and $D_{1+}$ denote outcome status between time 0 and 1 and after time 1, respectively. $U_0$ and $U_1$ denote status of another cause of the outcome at time 0 and 1, respectively.

*Immortal time arises from using postbaseline information to define exposure*

Suppose time zero (or baseline) was set as the day when the discovery was published. Nobel Prize winners have to survive until they have won their first award to be classified as winners; however, there is no such requirement for controls. Immortal time refers to the time between the publication of the discovery and the conferment of the Nobel Prize for winners (Figure 1a), which arises from using postbaseline information to define the exposure.

The bias generated is depicted in Figure 1b. The arrow from $E_0$ to $E_1$ means scientists who won Nobel Prizes at time 0 were also Nobel Prize winners at time 1. The arrow from $D_{0+}$ to $D_{1+}$ means scientists who died between time 0 and 1 were also dead after time 1. The arrow from $D_{0+}$ to $E_1$ means scientists have to be alive from time 0 to 1 to be classified as Nobel Prize winners at time 1. The structure of this bias is confounding, that is the presence of a common cause ($D_{0+}$) of the exposure ($E_1$) and the outcome ($D_{1+}$). It creates an open path between $E_1$ and $D_{1+}$, which can bias the association towards the direction favouring the winners. Specifically, winners having to remain alive until the Nobel Prize is awarded means survival confounds Nobel Prize winning on lifespan. Similarly, in a pharmacoepidemiologic study which uses postbaseline information to define exposure (e.g., at least one treatment during the follow-up), individuals have to be healthy enough to remain alive after recruitment until they receive the treatment, which confounds the effect of treatment on health outcomes. As such, the fundamental issue is differences between the individuals who do and do not survive the wait for treatment rather than counting time incorrectly.

***Immortal time arises from using postbaseline information to define eligibility***

Suppose time zero was set as the day when the discovery was published. All scientists who died before 75 years of age were excluded from the analysis. Immortal time refers to the time between the publication of the discovery and 75 years of age for both Nobel Prize winners and controls (Figure 1c), which arises from using postbaseline information to define the eligibility.

The bias generated is depicted in Figure 1d. The box around $D_{0+}$ means the analysis was restricted to scientists who remained alive until time 1. The structure of this bias is selection bias, that is conditioning on a common consequence ($D_{0+}$) of the exposure ($E_0$) and another cause of the outcome ($U_0$). Although conditioning on $D_{0+}$ closes the open path between $E_1$ to $D_{1+}$, it creates another open path between $E_0$ and $U_0$. If winning a Nobel Prize truly provides survival benefit, Nobel Prize winners who survive until 75 years of age are more likely to have another cause of death than controls who survive until 75 years of age without winning a Nobel Prize. Therefore, it can bias the association towards the opposite direction of the true effect. The exception is when the exposure, winning a Nobel Prize, has no causal effect on the outcome, survival; then $D_{0+}$ is not a collider (Supplemental Figure S1). Specifically, selecting on survival to 75 years when previous Nobel Prize status and other factors affect survival creates the classic *M*-bias, here specifically butterfly bias[21] given survival to 75 years affects current Nobel prize status and subsequent survival (Figure 1d). The magnitude of *M*-bias is generally smaller than confounding bias.[21]

Similarly, in a pharmacoepidemiologic study which uses postbaseline information to define eligibility (e.g., survival to one year after recruitment), selecting on survival to one year after recruitment when prior treatment and other factors affect survival also creates selection bias. Again, the fundamental issue is differences between the individuals who do and do not survive until eligibility rather than counting time incorrectly.

**Excluding immortal time from the follow-up does not fully address the bias**

Immortal time can be excluded from the follow-up by redefining time zero to a later timepoint, for some or all individuals in the study. We further illustrate the structure of bias in study designs which exclude immortal time from the follow-up.

*When immortal time arises from using postbaseline information to define exposure*

Suppose time zero was set as the day when Nobel Prize winners won their first award for winners, but as the day when the discovery was published for controls. Immortal time between the publication of the discovery and the first award for the Nobel Prize winner is thereby excluded from the follow-up (Figure 2a).

The DAG of this study design is depicted in Figure 2b, where S denotes excluded time. The arrow from $D_{0+}$ to $E_1$ persists because scientists still have to be alive from time 0 to 1 to be classified as Nobel Prize winners at time 1. The arrows from $E_0$ to S and from $E_1$ to S mean the excluded time (S) is determined by exposure status at time 0 and 1. For example, if a scientist won the first award at time 1 ($E_0 = 0$ and $E_1 = 1$), the person-time between time 0 and 1 would be excluded. The box around S means the analysis was restricted to unexcluded person-time. The structure of this bias is a composite of confounding and selection bias, that is the presence of a common cause ($D_{0+}$) of the exposure ($E_1$) and the outcome ($D_{1+}$) and conditioning on a common consequence (S) of the exposure ($E_0$) and the outcome ($D_{0+}$). Although it partially eliminates the guaranteed survival advantage of winners, it does not close the open path between $E_1$ and $D_{1+}$ and creates another open path between $E_0$ and $D_{0+}$.

*When immortal time arises from using postbaseline information to define eligibility*

Suppose time zero was set as 75 years of age. All scientists who died before 75 years of age were excluded from the analysis. Immortal time between the publication of the discovery and 75 years of age for both Nobel Prize winners and controls is excluded from the follow-up (Figure 2c).

The DAG of this study design is depicted in Figure 2d, where S denotes excluded time. The arrow from $D_{0+}$ to S means the excluded time (S) is determined by outcome status between time 0 and 1. For example, if a scientist died between time 0 and 1 ($D_{0+} = 1$), the person-time between time 0 and death would be excluded. The box around $D_{0+}$ means the analysis was restricted to scientists who remained alive until time 1. The box around S means the analysis was restricted to unexcluded person-time. The structure of this bias is selection bias, that is conditioning on a common consequence ($D_{0+}$) of the exposure ($E_0$) and another cause of the outcome ($U_0$). It is essentially the same as Figure 1d because conditioning on S does not generate a new open path.

**Competing risks can worsen bias from immortal time**

A competing risk is an event which precludes the occurrence of the outcome or alters the probability of the occurrence of the outcome.[22] There is no competing risk when the outcome is all-cause mortality; however, competing risks should be considered for all other outcomes. Consider another study to investigate the association of winning a Nobel Prize with the risk of dementia, where cardiovascular death is a competing risk. Nobel Prize winners and controls were identified as above.

In time-varying DAGs, $E_0$ and $E_1$ are exposure status at time 0 and 1, respectively. $D_{0+}$ and $D_{1+}$ are outcome status between time 0 and 1 and after time 1, respectively. $CR_{0+}$ and $CR_{1+}$ are status of a competing risk between time 0 and 1 and after time 1, respectively. $U_0$

and $U_1$ are status of a common cause of the outcome and the competing risk at time 0 and 1, respectively.

*When immortal time arises from using postbaseline information to define exposure*

Suppose time zero was set as the day when the discovery was published (Figure 3a). Figure 3b shows the bias with the presence of a competing risk. The arrows from $D_{0+}$ to $E_1$ and from $CR_{0+}$ to $E_1$ mean scientists have to be alive without the occurrence of dementia from time 0 to 1 to be classified as Nobel Prize winners at time 1. These arrows create open paths between $E_1$ and $D_{1+}$ and between $E_1$ and $CR_{1+}$. As $U_1$ creates another open path between $D_{1+}$ and $CR_{1+}$, an additional open path between $E_1$ and $D_{1+}$ is generated.

*When immortal time arises from using postbaseline information to define eligibility*

Suppose time zero was set as the day when the discovery was published, and all scientists who had a diagnosis of dementia or died before 75 years of age were excluded from the analysis (Figure 3c). Figure 3d shows the bias with the presence of a competing risk. The boxes around $D_{0+}$ and $CR_{0+}$ mean the analysis was restricted to scientists who remained alive without the occurrence of dementia until time 1. Although these boxes close the open paths between $E_1$ and $D_{1+}$ and between $E_1$ and $CR_{1+}$, they create two open paths between $E_0$ and $U_0$. The exception is when the exposure, winning a Nobel Prize, has no causal effect on either the outcome, dementia, or the competing risk, cardiovascular death; then neither $D_{0+}$ nor $CR_{0+}$ is a collider (Supplemental Figure S2).

**Bias from immortal time in pharmacoepidemiology**

Bias from immortal time is common in pharmacoepidemiologic studies.[4, 7] The bias has been named as "immortal time bias",[4, 7, 12] "survivor treatment selection bias",[23] "survivor bias",[24] or generally as "time-dependent bias" or "time-related bias".[6, 8, 9] The discrepancy in terms might have introduced difficulties in understanding the structures of bias arising from immortal time. Therefore, we emphasize the causal structure rather than nomenclature to appreciate the bias and inform design of epidemiologic studies. Table 1 summarizes the structures and sources of bias from immortal time in terms of confounding, selection bias or a combination of both.

Classical immortal time bias refers to the bias arising from using postbaseline information to define exposure, with or without the exclusion of immortal time (Figures 1b and 2b, respectively). Suissa has reviewed five cohort study designs leading to this bias.[4] Specifically, these studies define the exposure based on the mean number or a minimum or maximum number of prescriptions after cohort entry.[4] Immortal time refers to the period from cohort entry until a certain number of prescriptions is given for individuals classified as exposed. Time-based, event-based and multiple-event-based cohorts include immortal time in the follow-up,[4] which share the structure in Figure 1b. Exposure-based and event-exposure-based cohorts exclude immortal time from the follow-up,[4] which share the structure in Figure 2b.

Bias from immortal time can also occur in case-control studies, because there is always an underlying cohort, either explicit or virtual, for each case-control study.[25, 26] The bias arises in a case-control study, where the exposure was defined based on any prescription during the observed period.[25] Immortal time refers to the period from the start of the observed period until the first prescription for individuals classified as exposed. This bias named "time-window bias",[25] also shares the structure in Figure 1b.

Previous studies have introduced approaches to deal with classical immortal time bias.[24, 27] However, study designs which use postbaseline information to define eligibility might reduce confounding at the expense of introducing selection bias. A study design excludes all individuals who die within exposure window (i.e., a period during which the prescription status is used to define the exposure) and sets time zero as the end of exposure window.[24] This design shares the structure in Figure 2d. Another study design, prescription time-distribution matching, assesses the time of prescription for exposed individuals, randomly selects a time from this set for each control, and excludes controls who die before the time selected.[27] Time zero is set as the time of prescription for exposed individuals and as the time randomly selected for controls.[27] The structure of this design is depicted in Supplemental Figure S3 (Figure 1d is modified by adding arrows from $E_0$ and $E_1$ to S (excluded time) and a box around S). Nevertheless, these structures are fundamentally the same as Figure 1d, because they do not create new open paths between exposure and outcome.

Again, competing risks should be considered for outcomes other than all-cause mortality, particularly in studies investigating late-onset diseases among patients or older people. For example, a study investigating the association of statin use on the risk of prostate cancer among patients with heart disease, has cardiovascular death as a competing risk that cannot be ignored. More competing risks should be considered in studies involving older people, because the probability of occurrence of disease usually increases by age.

**Solutions to address bias from immortal time**

Bias from immortal time may be avoided by aligning time zero, exposure allocation, and eligibility for study inclusion,[28] and by excluding individuals with prior exposure.[29] To emulate a target trial, time zero should be set as the time when eligible individuals who have not been exposed before are classified into either group.[28] All confounders and selection bias (e.g., due to common causes of exposure/outcome and survival to recruitment) need to be adequately addressed to ensure exchangeability between groups at baseline, although this is not always feasible.[30] An analogue of the intention-to-treat analysis does not use postbaseline information to define exposure or eligibility, which removes the arrow from $D_{0+}$ to $E_1$ and the box around $D_{0+}$ and thereby mitigates the biases. In practice, the analogue of the intention-to-treat effect estimates could be uninformative when few exposures start at time zero.

Statistical approaches which focus on handling follow-up times correctly can be considered as an analogue of the per-protocol analysis, such as the person-time approach,[7] time-dependent analysis,[24] and the sequential approach.[27] The person-time approach accounts person-time before the start of exposure as unexposed and after that as exposed.[7] Time-dependent analysis codes the exposure status as a time-varying variable that changes from 0 to 1 when the exposure starts.[24] The sequential approach emulates a sequence of mini trials with increasing time zero when eligible individuals are classified into either group; individuals in the unexposed group are artificially censored when they are exposed.[27] Again, all confounders and selection bias should be fully addressed at baseline. In particular, the analogue of the per-protocol analysis should additionally adjust for time-varying confounders that affect change in exposure status/artificial censoring and the risk of the outcome.[31]

**Conclusion**

We illustrate that immortal time arises from using postbaseline information to define exposure or eligibility. We use time-varying DAGs to show that the structures of bias from immortal time are confounding by survival until exposure allocation or selection bias from selecting on survival until eligibility, but not measurement error of follow-up times. We further explain that excluding immortal time from the follow-up does not fully address the bias and that the presence of competing risks can worsen the bias. Epidemiologic studies should be designed and analysed to avoid or mitigate bias from immortal time.

**Declarations**

*Ethics approval*

This study does not require ethical approval.

*Data availability*

This study does not include original data.

*Supplementary data*

Supplementary data are available at *IJE* online.

*Author contributions*

GY and CMS contributed to the study conception. GY drafted the manuscript with critical feedback and revisions from SB and CMS. All authors read and approved the final version of the manuscript.

*Funding*

SB is supported by the Wellcome Trust (225790/Z/22/Z) and the United Kingdom Research and Innovation Medical Research Council (MC_UU_00002/7). The funders had no role in this study.

*Conflict of interest*

None declared.

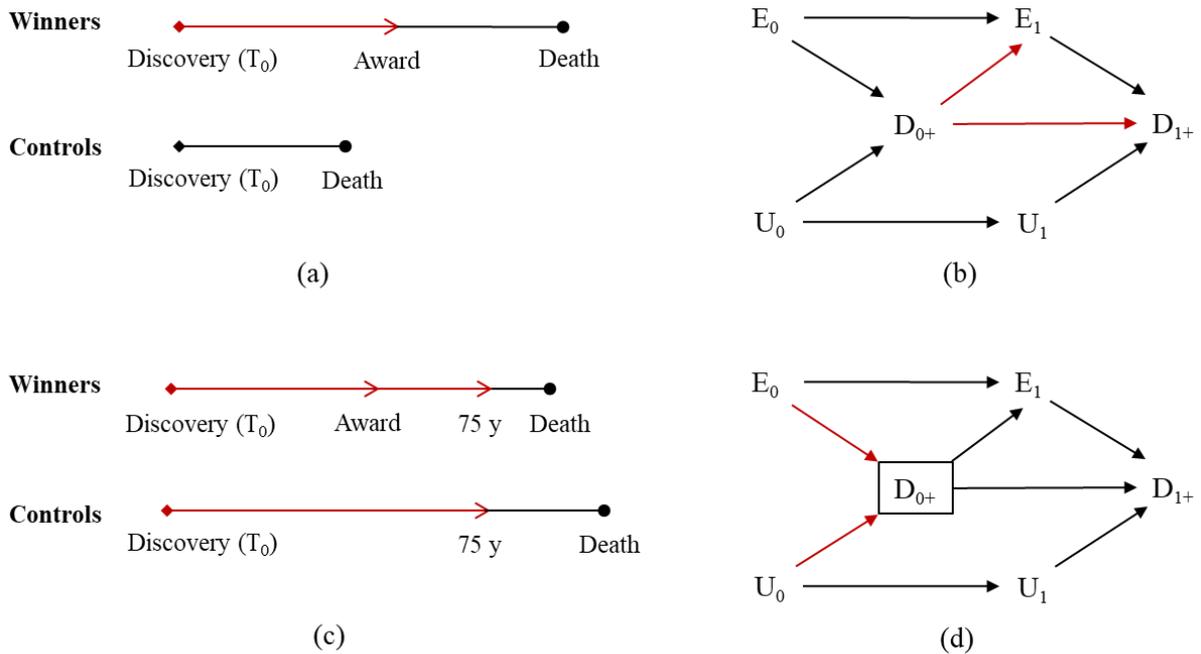

Figure 1. Illustrations and directed acyclic graphs (DAGs) of study designs with immortal time.

$T_0$ denotes time zero (or baseline) and red line denotes immortal time. $E_0$ and $E_1$ are exposure status at time 0 and 1, respectively. $D_{0+}$ and $D_{1+}$ are outcome status between time 0 and 1 and after time 1, respectively. $U_0$ and $U_1$ are status of another cause of the outcome at time 0 and 1, respectively. Red arrows denote key arrows that create open paths and result in bias.

(a) Time zero was set as the day when the discovery was published;

(b) Immortal time arises from using postbaseline information to define exposure;

(c) Time zero was set as the day when the discovery was published and all scientists who died before 75 years of age were excluded;

(d) Immortal time arises from using postbaseline information to define eligibility.

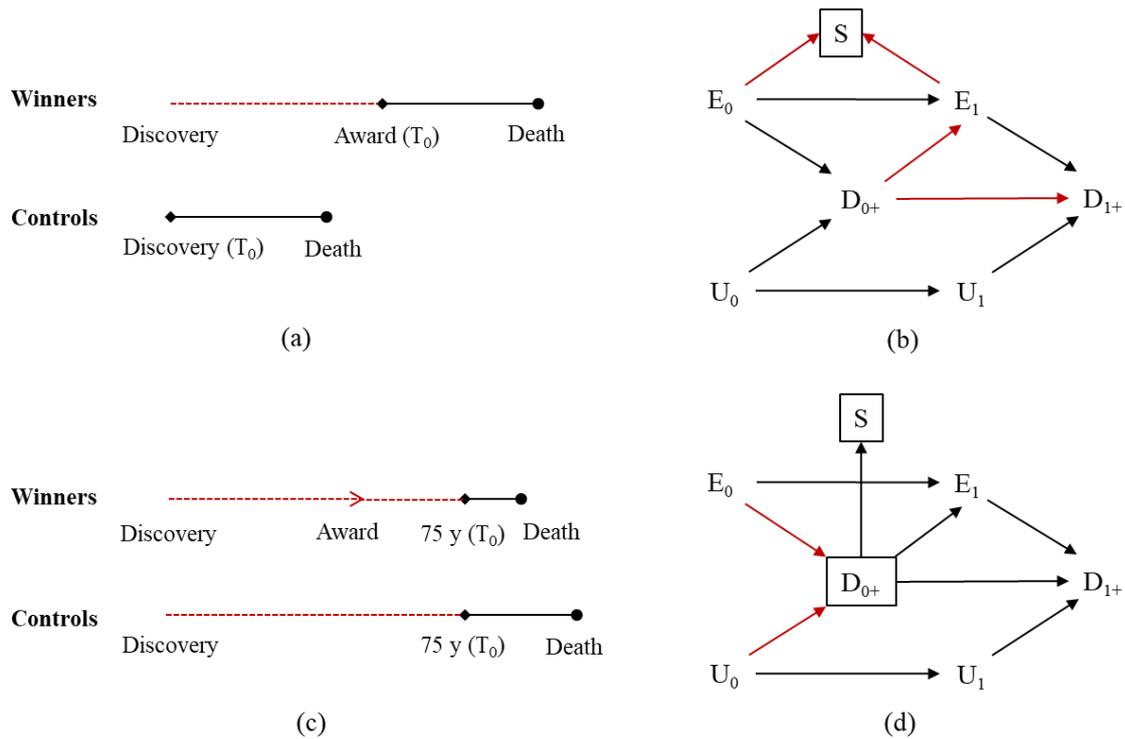

Figure 2. Illustrations and directed acyclic graphs (DAGs) of study designs excluding immortal time from the follow-up.

$T_0$ denotes time zero (or baseline) and red dotted line denotes immortal time excluded from the follow-up. $E_0$ and $E_1$ are exposure status at time 0 and 1, respectively. $D_{0+}$ and $D_{1+}$ are outcome status between time 0 and 1 and after time 1, respectively. $U_0$ and $U_1$ are status of another cause of the outcome at time 0 and 1, respectively. S is excluded time. Red arrows denote key arrows that create open paths and result in bias.

(a) Time zero was set as the day when Nobel Prize winners won their first award for winners, but as the day when the discovery was published for controls;

(b) Immortal time arising from using postbaseline information to define exposure is excluded;

(c) Time zero was set as 75 years of age and all scientists who died before 75 years of age were excluded;

(d) Immortal time arising from using postbaseline information to define eligibility is excluded.

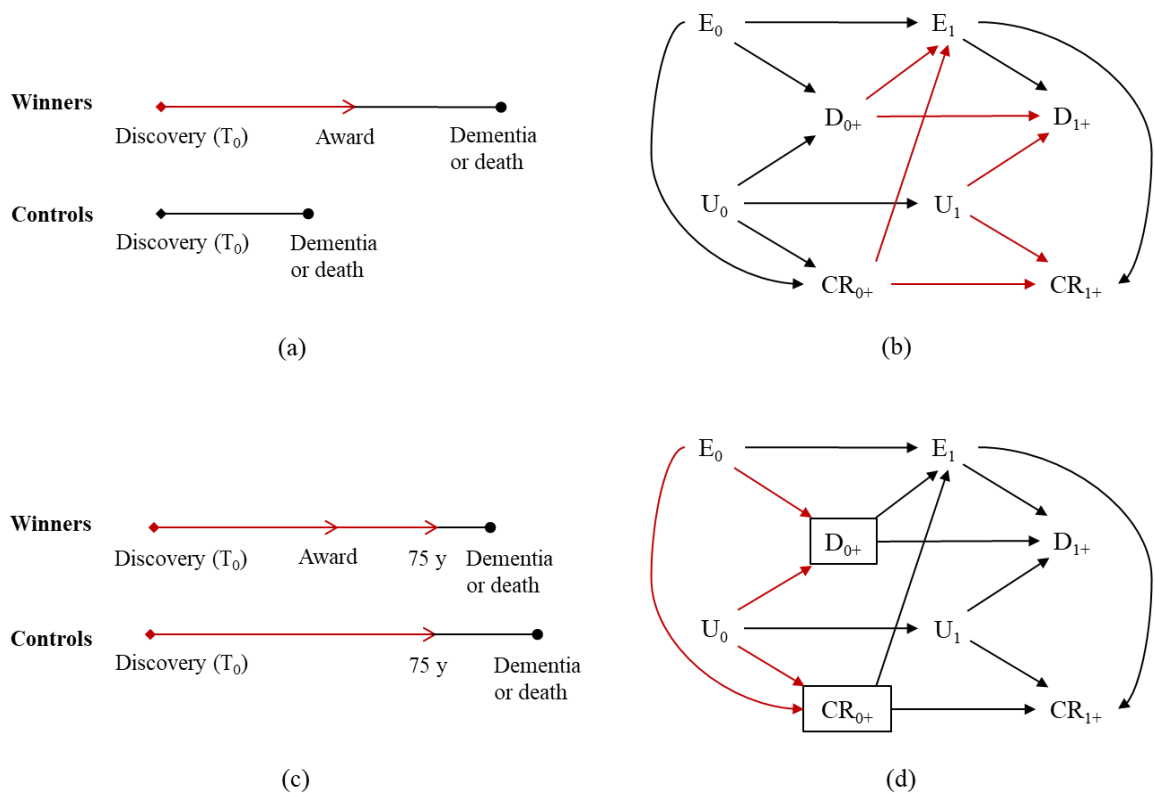

Figure 3. Illustrations and directed acyclic graphs (DAGs) of study designs with immortal time and the presence of a competing risk.

$T_0$ denotes time zero (or baseline) and red line denotes immortal time. $E_0$ and $E_1$ are exposure status at time 0 and 1, respectively. $D_{0+}$ and $D_{1+}$ are outcome status between time 0 and 1 and after time 1, respectively. $CR_{0+}$ and $CR_{1+}$ are status of a competing risk between time 0 and 1 and after time 1, respectively. $U_0$ and $U_1$ are a common cause of the outcome and the competing risk at time 0 and 1, respectively. Red arrows denote key arrows that create open paths and result in bias.

(a) Time zero was set as the day when the discovery was published;

(b) Immortal time arises from using postbaseline information to define exposure;

(c) Time zero was set as the day when the discovery was published and all scientists who had a diagnosis of dementia or died before 75 years of age were excluded;

(d) Immortal time arises from using postbaseline information to define eligibility.

Table 1. A summary of the structures and sources of bias from immortal time.

| Cause of immortal time | Exclusion of immortal time | Presence of competing risks [a] | Structure and source of bias from immortal time | Directed acyclic graphs |
|---|---|---|---|---|
| Define exposure by postbaseline information | No | No | Confounding by survival until exposure allocation | Figure 1b |
| Define eligibility by postbaseline information | No | No | Selection bias from selecting on survival until eligibility | Figure 1d |
| Define exposure by postbaseline information | Yes | No | A composite of confounding by survival until exposure allocation and selection bias by excluding immortal time | Figure 2b |
| Define eligibility by postbaseline information | Yes | No | Selection bias from selecting on survival until eligibility | Figure 2d |
| Define exposure by postbaseline information | No | Yes | Confounding by survival without occurrence of the outcome until exposure allocation | Figure 3b |
| Define eligibility by postbaseline information | No | Yes | Selection bias by selecting on survival without occurrence of the outcome until eligibility | Figure 3d |

[a] There is no competing risk when the outcome is all-cause mortality; however, competing risks should be considered for all other outcomes.

# Supplementary Material

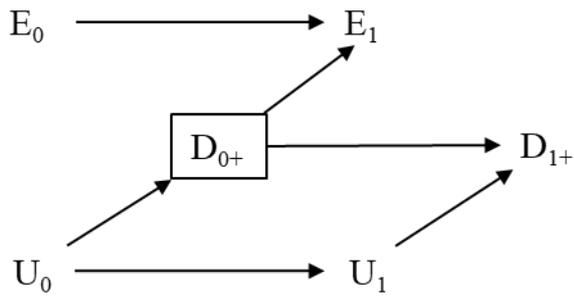

Supplemental Figure S1. Directed acyclic graphs (DAGs) of the study design with immortal time, where immortal time arises from using postbaseline information to define eligibility and the exposure has no causal effect on the outcome.

$E_0$ and $E_1$ are exposure status at time 0 and 1, respectively. $D_{0+}$ and $D_{1+}$ are outcome status between time 0 and 1 and after time 1, respectively. $U_0$ and $U_1$ are status of another cause of the outcome at time 0 and 1, respectively.

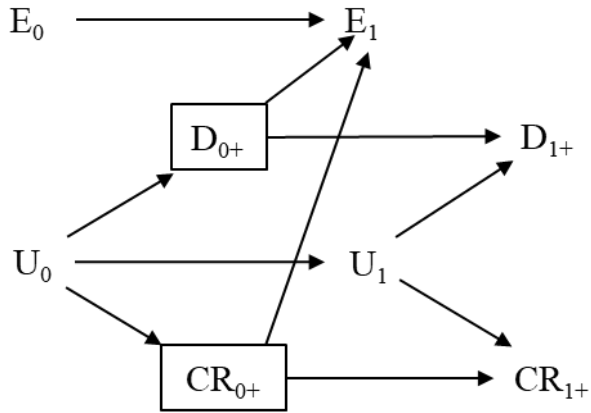

Supplemental Figure S2. Directed acyclic graphs (DAGs) of the study design with immortal time and the presence of a competing risk, where immortal time arises from using postbaseline information to define eligibility and the exposure has no causal effect on the outcome or the competing risk.

$E_0$ and $E_1$ are exposure status at time 0 and 1, respectively. $D_{0+}$ and $D_{1+}$ are outcome status between time 0 and 1 and after time 1, respectively. $CR_{0+}$ and $CR_{1+}$ are status of a competing risk between time 0 and 1 and after time 1, respectively. $U_0$ and $U_1$ are a common cause of the outcome and the competing risk at time 0 and 1, respectively.

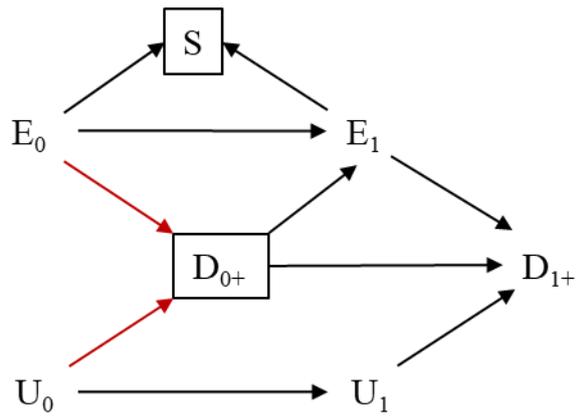

Supplemental Figure S3. Directed acyclic graphs (DAGs) of prescription time-distribution matching.

$E_0$ and $E_1$ are exposure status at time 0 and 1, respectively. $D_{0+}$ and $D_{1+}$ are outcome status between time 0 and 1 and after time 1, respectively. $U_0$ and $U_1$ are status of another cause of the outcome at time 0 and 1, respectively. S is excluded time. Red arrows denote key arrows that create an open path and result in bias.